# Different Applications and Technologies of Internet of Things (IoT)

Feisal Hadi Masmali [1,2], Shah J. Miah [3], Nasimul Noman[4]

[1] School of Electrical Engineering and Computing, The University of Newcastle, New South Wales, Australia
[2] College of Business Administration, Jazan University, Saudi Arabia
feisal.masmali@uon.edu.au
[3] Newcastle Business School, The university of Newcastle, Newcastle City Campus, New South Wales, Australia
shah.miah@newcastle.edu.au

[4] School of Information and Physical Sciences, University of Newcastle, NSW, Australia
nasimul.noman@newcastle.edu.au

**Abstract.** Internet of things (IoT) has significantly altered the traditional lifestyle to a highly technologically advanced society. Some of the significant transformations that have been achieved through IoT are smart homes, smart transportation, smart city, and control of pollution. A considerable number of studies have been conducted and continues to be done to increase the use of technology through IoT. Furthermore, the research about IoT has not been done fully in improving the application of technology through IoT. Besides, IoT experiences several problems that need to be considered in order to get the full capability of IoT in changing society. This research paper addresses the key applications of IoT, the architecture of IoT, and the key issues affecting IoT. In addition, the paper highlights how big data analytics is essential in improving the effectiveness of IoT in various applications within society.

## 1      Introduction

The term "Internet of Things" (IoT) is used to illustrate a network of physical objects, that is, "things" that are implanted with software, sensors as well as other technologies to facilitate connectivity as well as the exchange of information with other systems as well as devices over the internet [1]. Such devices range from objects in households to complex industrial tools. According to Rangers [1], there are above 7 billion connected IoT gadgets in the world, and the number is approximated to rise to 22 billion by 2025. As an upcoming paradigm, IoT allows communication between electronic devices and sensors via the internet, which significantly facilitates people's lives. In addition, IoT applies smart devices as well as the internet to develop solutions to challenges as well as issues that various businesses and private as well as public industries face across the world [2]. Furthermore, IoT has continuously turned out to be a vital aspect of our lives, which is evident everywhere in the world. In general, IoT integrates different varieties of smart systems frameworks as well as intelligent devices as well as sensors [2]. Besides, IoT exploits quantum as well as nanotechnology in the form of sensing, storage as well as processing speed, which was not attainable earlier [3]. There have been extensive studies that have been conducted in order to illustrate the effectiveness as well as the applicability of IoT transformations. This paper provides an in-depth illustration of IoT in terms of its architecture, its major applications, significant problems and issues of IoT, and the significance of big data analytics within IoT.

There is a significant transformation that is evident from our daily lives whereby the application of IoT devices as well as technology has significantly increased. The establishment of Smart Home Systems (SHS), home automation systems, and appliances that use devices that rely on the internet are some of IoT developments [4]. Another significant gain of IoT is the Smart Health Sensing System (SHSS). This system integrates several small intelligent pieces of equipment as well as devices that assist in supporting health. The system is also applied to inspect critical health conditions within the hospitals. This signifies that IoT has altered the medical domain scenario by introducing convenient technology as well as smart devices [5]. Furthermore, the developers, as well as researchers of IoT, have actively been participating in improving the lifestyle of both disabled and aged individuals, whereby IoT has an extreme performance within this sector. IoT has indicated a new direction for the ordinary life of the aged and disabled individuals through the development of devices that are cost-effective and easy to use for these individuals [2].

The other important aspect of IoT in daily life is transportation. It has contributed to new advancements making transportation more efficient, comfortable as well as reliable. For instance, intelligent sensors and drone devices have been widely applied in controlling traffic across major cities. Additionally, there are new vehicles that consist of sensing devices that allow sensing heavy traffic congestions and to find alternative routes on the map with low traffic congestion [6]. This clearly indicates that IoT has played a significant role in different aspects of life as well as technology.

Besides, IoT has displayed its significance as well as potential within the economic and industrial fields. It has been regarded as a revolutionary tread within the trade and stock exchange market. Despite the potential associated with IoT in life improvement, the issue of security of information is a crucial concern, which is the main challenge faced in IoT [5]. Hackers have actively been using the internet in conducting cyber-attacks, which has made data insecure. Despite this, IoT developers and researchers have been involved in giving the best possible solutions to cope with data security issues.

## 2      The IoT System Architecture

The architecture of an IoT system is typically illustrated as a process that involves steps whereby information flows from sensors incorporated in "things" via a network, and finally onto a corporate data center or the cloud whereby it undergoes processing, analysis as well as storage [7]. In this case, a "thing" may be a machine, a person, or a building. In the IoT system process, also send information to the other direction instructing the actuator or the connected devices to take various actions in order to control a physical process.

There are four essential layers incorporated in an IoT architecture that describe the workability of an IoT System. They are the perception layer, network layer, application layer, and business layer [2]. The perception is the lowest in the IoT structure. It constitutes physical devices such as sensors, Radio-frequency identification chips and barcodes, which are linked to the IoT system. They assist in gathering data that is then delivered to the network layer.

Moreover, the network layer functions as a transmission medium of data from the perception layer to the data processing system [2]. This channelling of data may apply either be through a wired or a wires medium. The application layer utilizes the processed data for global device management [2]. The business layer is found on the top of the IoT architecture. It usually controls the entire IoT system, that is, its applications as well as services [2]. It visualizes data from the application layer and uses the information to generate plans as well as strategies. According to Kumar et al. [2], the IoT architecture can be altered according to the need as well as the application domain.

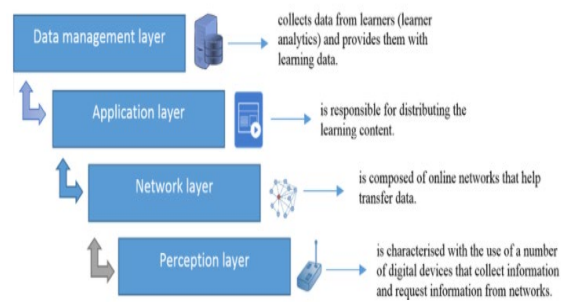

**Fig. 1.** The architecture of IoT System (Source: Jahnke A. (2020). "The 4 Stages of IoT Architecture.")

Apart from layered frameworks, IoT systems contain other functional blocks that help in various IoT services like sensing methods, authentication as well as identification, control, and management [8]. Besides, these functional blocks are involved in connectivity issues, audio or video monitoring, input or output operation, and management of storage. These blocks all form an effective IoT system essential for maximum performance. According to Nicolescu et al. [9], several architecture references with technical specifications that have been proposed still deviate from the typical IoT architecture accepted for global IoT.

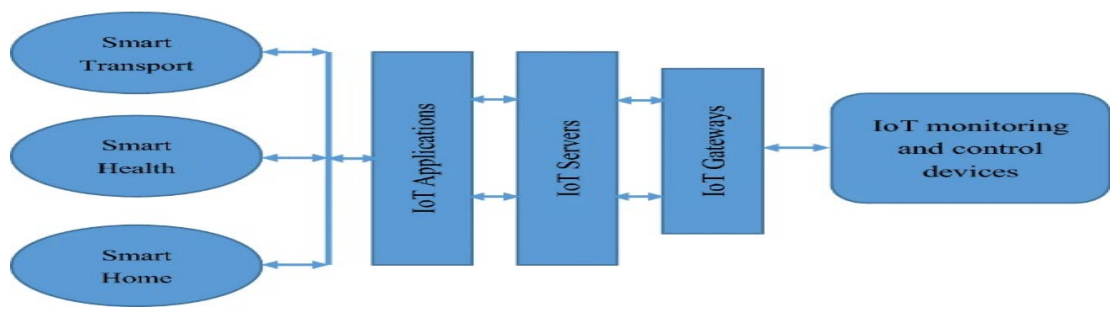

**Fig. 2.** The generic structure of IoT (Source: Kumar, S., Tiwari, P., & Zymbler, M. (2019). Internet of Things is a revolutionary approach for future technology enhancement: a review. Journal of Big Data, 6(1)).

According to Kumar et al. [2], the critical design factors for a productive IoT architecture within a heterogeneous condition are scalability, openness, modularity, and interoperability. The design must include an objective of meeting the needs of cross-domain interactions, big data analytics, storage and applications that are user-friendly. The IoT structure must also have the ability to scale up the workability and include intelligence as well as automation within the IoT devices [2].

Furthermore, the increased amount of data generated from IoT sensors and IoT gadgets has contributed to a big issue. This implies that an effective IoT structure should be able to cope with the increased amount of data within the IoT system. The leading IoT system architectures are fog and cloud computing. They are responsible for handling, monitoring as well as analysis of a large amount of information within IoT systems [2].

Sensors, as well as actuators, play a vital role in stage one of the IoT architecture. Sensors have the capacity to detect data flow from the real world, which constitutes things like humans, vehicles, animals, electronic devices, and buildings. The sensors transmit this data, which may be used further for analysis. On the other hand, actuators are capable of intervening the reality, for instance, turning off the music, slowing down the speed of a vehicle, and regulating temperature in a room. In general, stage one of the IoT architecture enables collecting information from the real world. This information may be applied for more analysis.

Moreover, stage two enables cooperation between sensors and actuators together with gateways as well as information acquisition systems. In addition, this stage allows aggregation and optimization of the enormous data gathered from stage one to appropriate way for the purpose of processing [2]. The data is then passed to stage three of the IoT architecture, which is edge computing. This sage can be described as a free architecture that enables the utilization of technologies as well as huge computing power from different areas in the world [2]. In addition, edge computing provides a robust way of streaming data processing, and therefore it is acceptable for IoT systems.

During stage three of IoT architecture, technologies of edge computing manage a huge amount of information and give several functionalities like imaging, data integration from different sources, and analysis of data through machine learning techniques. The final stage involves various crucial activities like deep analysis and processing of information, giving feedback to increase the whole system's precision. All processes in this stage are conducted in the cloud server [2]. Moreover, this stage involves big data frameworks like Spark in order to handle substantial streaming data. Furthermore, machine-learning methods may be applied to establish suitable prediction models that may help develop a reliable and accurate IoT system that meets the requirements of people [2].

## 3       Major Applications of IoT

IoT has a wide range of applications since it can be adjusted to almost every technology that can provide suitable information concerning its operation, the fulfilment of an activity, and concerning the environmental situations that require to be monitored and regulated distantly.  A significant number of industries are currently adopting IoT technologies to simplify, enhance, and automate as well as to control distinct processes.

## 4       Smart City, Transport As Well As Vehicles

IoT has significantly altered the conventional civil structure of society to form a high technological structure that involves concepts of smart city, smart home as well as smart vehicles. There are vast advancements that are aided by supporting technologies like machine learning [10]. Furthermore, several technologies like cloud server technology as well as wireless sensor networks and IoT servers must be applied jointly in order to generate an efficient smart city. However, it is vital to consider the environmental factor of the smart city. In connection to this, there should be the establishment of energy-efficient technologies as well as green technologies when it comes to the establishment of smart city infrastructure.

Furthermore, new vehicles are currently incorporated with smart devices, which can easily detect traffic congestion on roads and helps in providing alternative routes. This significantly helps in minimizing congestions n within the city. Besides, there should be a design of smart devices with optimum cost that can be compatible with all vehicles in monitoring different parts of the vehicle.  In addition, IoT significantly helps in maintaining the condition of a vehicle. Also, self-driving vehicles have the capacity to pass information to other vehicles via intelligent sensors. This allows smooth flow of the traffic compared to human-driven vehicles.  Although this technology will require significant time to spread worldwide, IoT gadgets will continue helping in sensing traffic congestion.

Sensors in fleet vehicles facilitate effective communication between the managers of the vehicle and the vehicle. The manager or the driver of the vehicle can easily get information concerning the status, operation as well as requirement of the vehicle by evaluating the software responsible for gathering, processing, and organizing information. In general, the integration of IoT to fleet management helps in geolocation and performance analysis and may allow the generation of relevant information that can help improve driving vehicles.

## 5       Maintenance Management

This forms one of the extensive deployment areas of IoT technology. The amalgamation of sensors and software specialized in Enterprise asset management (EAM) as well as computerized maintenance management systems (CMMS) allowed the development of a multifunctional tool that can be applied in various disciplines as well as practices. The main objective of these tools is to increase the productive life of assets. The application of these tools is almost unlimited since the software attributes that process and organize data gathered by sensors are specifically designed to address the management requirements of physical assets [11].

## 5       Healthcare

According to Kumar, Tiwari and Zymbler[2], IoT is entirely committed to providing financial benefits as well as development to society. These developments include but are not limited to public facilities such as water and quality maintenance, economic development, and industrialization. When it comes to the health sector, the application of wearable sensors that are linked to patients enables healthcare providers to observe a patient's condition while outside the healthcare facility or in real-time. Via progressive inspection of various metrics as well as automatic alerts about their crucial signs, the IoT assists in enhancing patients care and prevention of lethal situations among high-risk patients [11]. Moreover, the integration of IoT in hospital beds is another essential application of IoT within the healthcare sector. This has given way to smart beds that are incorporated with sensors to monitor patients' vital signs such as blood pressure and body temperature.

## 6    Wearables Devices

These are small devices that do not consume a lot of energy and are equipped with sensor and are equipped with the relevant hardware to take measurements as well as readings and their software gather and organize data concerning the users. Some of the examples of wearable devices that have been widely applied are virtual glasses, GPS tracking belts, and fitness bands to observe, for instance, calories use. Corporations like Google and Samsung have established and are using IoT in their devices [11].

## 7    Hospitality

IoT integration within the hotel industry enables improvement in the provision of services. For instance, the implementation of electronic keys that are directly sent into guests' mobile devices allows the automation of several interactions. Some of the crucial activities that can be easily managed through the application of IoT in the hotel industry are the realization of orders or services in a room and the automated charging of room [11]. Moreover, the process of checking out is automated with the help of electronic keys.

## 8    Smart Grid as Well as Energy Saving

IoT has enabled better monitoring as well as control of the electrical network. This has been effectively achieved by the continuous application of smart meters with sensors, and installing sensors at various strategic locations from the point of power production to distribution areas. These sensors significantly allow bidirectional communication between the power companies and users, whereby essential information pertaining to faults, repairs, or decision-making can be obtained. Moreover, IoT allows sharing critical information to the end-users concerning their power consumption patterns and concerning the best way of reducing their expenditure on power [11].

## 9    Water Supply

Moreover, IoT has progressively been applied in the supply of water. A sensor that is equipped internally or externally in water meters, connected to the internet, and integrated with the relevant software assist in collecting, processing, and analysing data. This information is essential as it allows detecting faults, understanding water consumers' behaviors, reporting the outcomes, and providing a course of action to the company. Similarly, the application of IoT in water supply allows the consumers to track information about their water consumption via a webpage and even in real-time, whereby they can get alerts when consumption is exceeded the average consumption.

## 10    Agriculture

Agriculture is a vital sector to the rapidly growing world's population. Therefore, there is a need for the advancement of present agricultural activities in order to satisfy the huge world's population. This calls for the amalgamation of agriculture with advanced technology in order to increase production levels. One of the probable methods is through greenhouse technology towards this direction. Greenhouses provide a way of controlling environmental parameters to increase production. However, this has been improved by automating processes through the application of IoT. Through smart devices as well as sensors, environmental parameters can be easily monitored. This contributes to improved production and energy saving.

## 11    Key Issues Associated with IoT

There are several issues associated with the inclusion of IoT in almost all aspects of daily life and technologies associated with the transmission of data between the IoT integrated devices. Besides, these issues have raised challenges to IoT developers and researchers with the advanced smart society. Moreover, as technology continues to progress, there is also a need to advance IoT systems.

## 12    Privacy and Security

One challenge facing IoT systems is security as well as privacy. This is due to the available threats, cyber-attacks as well as vulnerabilities [12]. Some of the factors that contribute to privacy issues on IoT embedded devices include inadequate authorization as well as authentication, web interface, firmware, insecure software, and faulty transport layer encryption [13]. According to Sisinni et al. [14], security, as well as privacy, are essential in confidence development in IoT Systems. In order to prevent attacks, security techniques should be incorporated at each layer of IoT [15]. Furthermore, various protocols have been established at each layer of communication in order to ensure that IoT-based systems are secure from threats [16]. An example of a security protocol that has been established between the transport and application layer in order to secure IoT systems is the Secure Socket Layer (SSL). Different IoT systems require varied methods in the provision of communication security solutions between IoT devices. Furthermore, wireless communication between IoT devices increases the risk of security issues.

On the other hand, privacy is a vital aspect of IoT as ill allows IoT devices users to feel safe while applying IoT systems. It is critical to maintaining authorization as well as authentication in a safe network to develop trust between parties communicating with each other [17]. Moreover, there are various privacy policies for distinct devices within the IoT systems.

To counter this issue, each object should have the capacity to verify other objects' privacy policies within the IoT system before transferring information.

## 13  Inter-Operation and Standard Issues

Inter-operability or integration refers to the practicability in the switching of data among several IoT devices as well as systems. Transmission of information does not necessarily depend on the devices and systems applied. Inter-operability issues emerge because of the heterogeneous nature of various technologies as well as solutions involved in the development of IoT. According to Interoperability in Digital Health [18], the levels of inter-operability are technical, semantic-syntactic as well as organizational. Moreover, IoT systems incorporate several functionalities in order to improve inter-operability, which improves communication among various IoT objects within a heterogeneous environment. Some of the suggested ways of handling inter-operability are adapters/gateways based and virtual networks/overlay based [19].

## 14  Ethics, Law and Rights

Ethics, legality, as well as regulatory rights are significant considerations for IoT developers. To safeguard morals and prevent people from breaching them, some specific policies and procedures should be considered. The sole difference that lies between ethics and legislation is that ethics are the guidelines that individuals are required to stick to, whereas legislations are the limitations pressed by the government. Moreover, ethics, as well as legislation, are intended to protect the quality and protect individuals from taking part in criminal behavior. Tzafestas [20] indicates that there are various real-world problems, which have been resolved due to the development of IoT, but at the same time, resulting in various ethical as well as legal problems. These problems include privacy, security of information, trust, and safety of data usability. Due insufficient trust with IoT, many users align with government law regarding protection, privacy, and information security. Therefore, it is vital to consider such issues in order to maintain and build up public trust in the deployment of IoT gadgets.

## 15  Dependability, Scalability and Accessibility

When other devices are included in a system without decreasing its effectiveness, the system is regarded as scalable. The fundamental drawback with IoT is incorporating many devices with varying memory, processor, storage capability, bandwidth accessibility, and reliability [21]. Access to resources forms another significant factor that should be considered. Within the IoT architecture layers, scalability, as well as availability, should be applied jointly. A good example of a system that portrays scalability is the Cloud-based IoT system. This is because they allow users to develop their IoT network by introducing other gadgets, storage, as well as processing capabilities as required.

The other important issue is making sure that resources to the suitable objects are accessible without necessarily bothering their location or time they are required. Several IoT networks are linked to global IoT systems in a scattered approach in order to maximize usage of their resources as well as services [2]. Various services, as well as resource availability, may be obstructed because of the availability of other different information transmission channels, like satellite communication. Therefore, an independent, as well as a dependable data transmission channel is required in order to ensure that resources and services are always available.

## 16  Quality of Service (QoS)

The other key issue in IoT systems is the quality of service (QoS) provided. QoS evaluates the quality, effectiveness, and productivity of IoT architecture, devices, and systems [22]. According to Huo and Wang [23], reliability, affordability, energy utilization, security, availability, and service time are significant in IoT systems. A good IoT platform should meet the necessities of QoS standards. Additionally, any IoT service should initially be defined to verify its dependability. People may also be free to articulate their requirements and needs accordingly. According to White, Nallur and Clarke [24], many techniques can be used to analyze QoS, but a trade-off exists between quality variables and approaches. In order to address this trade-off, high-quality models must be used.

A few high-quality models like OASIS-WSQM may be applied in assessing the methodologies used for QoS evaluation [25]. Such models include a large number of quality parameters that are enough for determining QoS for IoT services.

## 17  Big Data Analytics in IoT

In an IoT system, many devices and sensors share information. In addition, the increased development of IoT networks has contributed to the increase of IoT devices and sensors at a rapid rate. There is, therefore, the transmission of an enormous amount of data by these sensors and devices through the internet. The data transmitted is massive and, therefore, can be regarded as big data. Moreover, the increase in IoT embedded networks has resulted in issues like management and gathering of information, storage as well as processing, and analytics [2].

In smart buildings, the IoT big data system assists in dealing with issues like oxygen level management, detection of hazardous gases, and luminosity [26]. These systems enable the collection of information from sensors within the building and conduct data analytics to come up with decisions. Furthermore, Lee, Yeung and Cheng [27], indicates that production in an industry can be raised by deploying an IoT embedded Cyber-physical platform that incorporates data analysis and knowledge gathering techniques.

Besides, traffic congestion within smart cities is a vital issue when it comes to big data analytics. Traffic light signals are able to gather real-time traffic data with the help of sensors and IoT devices incorporated in them. The information gathered can be easily analyzed within an IoT-based traffic management system [2].

Moreover, in the healthcare sector, IoT sensors help in generating more information concerning patient health. The huge amount of data gathered requires to be combined into one database for quick processing in order to enable developing accurate decisions. Big data technology effectively provides a solution for this [28]. According to Mourtzis, Vlachou and Milas [29], when IoT is combined with big data analytics, it can effectively assist in changing the traditional methods applied in manufacturing industries. Devices that use sensors produce data that can be analyzed via big data techniques, which aid in making various decisions.

Besides, the application of cloud computing, as well as for analytics, is of great significance to energy development as well as conservation due to decreased costs as well as customer satisfaction [30]. In addition, IoT devices usually produce a large amount of information that requires an effective storage mechanism for further analysis in order to develop accurate real-time decisions. To deal with this huge data, Deep learning assists by providing high precision [31]. Therefore, Deep learning, IoT as well as big data analytics are all essential in creating a technologically advanced society.

## 18 Conclusion

A significant transformation is evident from our daily lives whereby the application of IoT devices as well as technology has significantly increased. Some of the major applications of IoT are smart homes, smart transportation, smart city, and control of pollution. Moreover, the current advancements in IoT have made researchers, as well as developers, collaborate in improving technology in IoT on a large scale, which will help society to the highest level. Furthermore, in order to achieve these improvements, we must consider various issues and challenges facing IoT, such as environmental issues. In addition, since IoT does not only provide services but also is involved in generating a large amount of information, it is essential to consider big data analytics since it helps in providing precise real-time decisions that could be applied in the establishment of an advanced IoT system.

Further studies can be conducted in multiple areas. For example, IoT functions can be combined to latest technologies such as block chain in healthcare development [32]. An application development study can be conducted through adopting the design science research (e.g. [33, 34, 35]). Example application with IoT can also extend the associative knowledge of Big Data Analytics solution design (e.g. in enhancing data sensing and collection practice) in tourism [36, 37, 38], education [39] and other associated problem domain.

## References

1. Ranger S. (2020). "What is the IoT? Everything you need to know about the Internet of Things right now."
2. Kumar, S., Tiwari, P., & Zymbler, M. (2019). "Internet of Things is a revolutionary approach for future technology enhancement: a review. *Journal Of Big Data*, 6(1). https://doi.org/10.1186/s40537-019-0268-2
3. Gatsis, K., & Pappas, G. (2017). "Wireless Control for the IoT." *Proceedings of the Second International Conference on Internet-Of-Things Design And Implementation*. https://doi.org/10.1145/3054977.3057313
4. Zhou, J., Cao, Z., Dong, X., & Vasilakos, A. (2017). "Security and Privacy for Cloud-Based IoT: Challenges." *IEEE Communications Magazine*, 55(1), 26-33. https://doi.org/10.1109/mcom.2017.1600363cm
5. Minoli, D., Sohraby, K., & Kouns, J. (2017). "IoT security (IoTSec) considerations, requirements, and architectures." *2017 14Th IEEE Annual Consumer Communications & Networking Conference (CCNC)*. https://doi.org/10.1109/ccnc.2017.7983271
6. Behrendt, F. (2019). "Cycling the Smart and Sustainable City: Analyzing EC Policy Documents on Internet of Things, Mobility and Transport, and Smart Cities." *Sustainability*, 11(3), 763. https://doi.org/10.3390/su11030763
7. Jahnke A. (2020). "The 4 Stages of IoT Architecture."
8. Causevic S., Colakovic A., Haskovic A. (2018). "The model of transport monitoring application based on the Internet of Things."
9. Nicolescu, R., Huth, M., Radanliev, P., & De Roure, D. (2018). Mapping the Values of IoT. *Journal of information technology, 33*(4), 345-360. https://doi.org/10.1057/s41265-018-0054-1
10. Park, E., del Pobil, A., & Kwon, S. (2018). "The Role of Internet of Things (IoT) in Smart Cities: Technology Roadmap-oriented Approaches." *Sustainability*, 10(5), 1388. https://doi.org/10.3390/su10051388
11. FRACTTAL (2019). "The 9 most important applications of the Internet of Things (IoT)."
12. Babovic, Z.B., Protic, V. and Milutinovic, V. (2016). "Web Performance Evaluation for Internet of Things Applications."
13. Home Security Systems. (n.d.). "HP News - HP Study Finds Alarming Vulnerabilities with Internet of Things (IoT)." Www.hp.com.
14. Sisinni, E., Saifullah, A., Han, S., Jennehag, U., & Gidlund, M. (2018). "Industrial Internet of Things: Challenges, Opportunities, and Directions." IEEE Transactions on Industrial Informatics, 14(11), 4724–4734. https://doi.org/10.1109/tii.2018.2852491

15. Yan, Z., Zhang, P., & Vasilakos, A. V. (2014). "A survey on trust management for Internet of Things." *Journal of Network and Computer Applications*, *42*, 120–134. https://doi.org/10.1016/j.jnca.2014.01.014
16. Pei, M., Cook, N., Yoo, M., Atyeo, A., & Tschofenig, H. (2016). *The Open Trust Protocol (OTrP)*. IETF.
17. Roman, R., Najera, P., & Lopez, J. (2011). "Securing the Internet of Things." *Computer*, *44*(9), 51–58. https://doi.org/10.1109/mc.2011.291
18. Interoperability in Digital Health (n.d.). Reference Material | Publications. Publications.iadb.org. https://publications.iadb.org/publications/english/document/Interoperability_in_Digital_Health_Reference_Material_en.pdf
19. Noura, M., Atiquazzaman, M. and Gaedke, M. (2019). *"Interoperability in Internet of Things Taxonomies and Open Challenges. Mobile Networks and Applications." 24, 796-809. - References - Scientific Research Publishing*. Www.scirp.org.
20. Tzafestas, S. G. (2018). Ethics and law in the internet of things world. *Smart cities*, *1*(1), 98-120.
21. Pereira, C., & Aguiar, A. (2014). "Towards Efficient Mobile M2M Communications: Survey and Open Challenges." *Sensors*, *14*(10), 19582–19608. https://doi.org/10.3390/s141019582
22. Nacera T., Abdelghani C., Karim D., and Mohamed Ahmed-Nacer (2018). "A Distributed Agent-Based Approach for Optimal QoS Selection in Web of Object Choreography."
23. Huo, L., & Wang, Z. (2016). "Service composition instantiation based on cross-modified artificial Bee Colony algorithm." *China Communications*, *13*(10), 233–244. https://doi.org/10.1109/cc.2016.7733047
24. White, G., Nallur, V., & Clarke, S. (2017). "Quality of service approaches in IoT: A systematic mapping." *Journal of Systems and Software*, *132*, 186–203. https://doi.org/10.1016/j.jss.2017.05.125
25. Oasis (2012). "Web Services Quality Factors Version 1.0 Candidate OASIS Standard 01."
26. Saggi, M. K., & Jain, S. (2018). "A survey towards an integration of big data analytics to big insights for value-creation." *Information Processing & Management*, *54*(5), 758–790. https://doi.org/10.1016/j.ipm.2018.01.010
27. Lee C. K. M., Yeung C.L. and Cheng M. (2015). "Research on IoT based Cyber-Physical System for Industrial big data Analytics." *In 2015 IEEE International Conference on Industrial Engineering and Engineering Management (IEEM)*, DOI: 10.1109/IEEM.2015.7385969
28. Vuppalapati C., Ilapakurti A. and Kedari S. (2016). "The Role of Big Data in Creating Sense EHR, an Integrated Approach to Create Next-Generation Mobile Sensor and Wearable Data-Driven Electronic Health Record (EHR)." *In the 2016 IEEE Second International Conference on Big Data Computing Service and Applications (BigDataService),* 10.1109/BigDataService.2016.18
29. Mourtzis D., Vlachou E. and Milas N. (2016). "Industrial Big Data as a Result of IoT Adoption in Manufacturing." *Procedia CIRP,* 55, 290-295
30. Ramamurthy, A., & Jain, P. (2017). *The Internet of Things in the Power Sector Opportunities in Asia and the Pacific ADB Sustainable Development Working Paper Series No. 48*. https://www.adb.org/sites/default/files/publication/350011/sdwp-48.pdf
31. Mohammadi, M., Al-Fuqaha, A., Sorour, S., & Guizani, M. (2018). "Deep learning for IoT big data and streaming analytics: A survey". *IEEE Communications Surveys & Tutorials*, *20*(4), 2923-2960.
32. Prokofieva, M. & Miah, S.J. (2019). "Blockchain in healthcare", *Australasian Journal of Information Systems*, 23, 1-22
33. Genemo, H., Miah, S.J. & McAndrew, A. (2015). "A Design Science Research Methodology for developing a Computer-Aided Assessment Approach using Method Marking Concept", Education and Information Technologies, 22, 1769–1784
34. Miah, S.J., Gammack, JG., & McKay, J. (2019). "A Metadesign Theory for Tailorable Decision Support", *Journal of the Association for Information Systems,* 20 (5), 570-603
35. Miah, S.J. & McKay, J. (2016). "A New Conceptualisation of Design Science Research for DSS Development, *Proceedings of the 20th Pacific Asia Conference on Information Systems (PACIS 2016)*, Taiwan, URL: http://www.pacis2016.org/Abstract/ALL/758.pdf
36. Miah, S.J., Vu, HQ. & Gammack, J. (2019). "A Big-Data Analytics Method for capturing visitor activities and flows: the case of an Island Country", *Information Technology and Management*, 20 (4), 203-221
37. Miah, S.J., Vu, HQ, & Gammack, J. (2018). "A Location Analytics Method for the Utilization of Geo-tagged Photos in Travel Marketing Decision-Making", *Journal of Information and Knowledge Management*, 18(1), 1-13
38. Miah, S.J., Hasan, N., & Gammack, J.G. (2020). "A Methodological Requirement for designing Healthcare Analytics Solution: A Literature Analysis", *Health Informatics Journal,* 26(4), 2300-2314
39. Miah, S.J., Miah, M., & Shen, J. (2020). "Learning Management Systems and Big Data Technologies for Higher Education", *Education and Info Technologies*, 25, 725–730